\documentclass[showpacs,twocolumn,floats,superscriptaddress]{revtex4}
\usepackage{graphicx}
\usepackage{amsmath}

\newcommand{\bq}{\begin{equation}}
\newcommand{\ee}{\end{equation}} \newcommand{\fr}[2]{\frac{#1}{#2}}
\newcommand{\eps}{\varepsilon}

\begin{document}

\title{Quantum dots in graphene}
\author{P.G. Silvestrov}
\affiliation{Theoretische Physik III,
Ruhr-Universit{\"a}t Bochum, 44780 Bochum, Germany}
\author{K.B. Efetov }
\affiliation{Theoretische Physik III,
Ruhr-Universit{\"a}t Bochum, 44780 Bochum, Germany}
\affiliation{L. D. Landau Institute for Theoretical Physics,
117940 Moscow, Russia}
\date{\today }

\begin{abstract}
We suggest a way of confining quasiparticles by an external
potential in a small region of a graphene strip. Transversal
electron motion plays a crucial role in this confinement.
Properties of thus obtained graphene quantum dots are investigated
theoretically for different types of the boundary conditions at
the edges of the strip. The (quasi)bound states exist in all
systems considered. At the same time, the dependence of the
conductance on the gate voltage carries an information about the
shape of the edges.
\end{abstract}

\pacs{73.63.Kv        73.63.-b        81.05.Uw }
\maketitle

Recently, single layers of carbon atoms (graphene) have been
obtained experimentally \cite{Novoselov04}. This new conducting
material with a high mobility \cite{Novosel05,Zhang05} has
attracted a lot of theoretical attention because of its special
band
structure~\cite{Fse,ando,Cheian06,Katsnel06,BreyF06,Tworzyd06}.
The spectrum of excitations in graphene consists of two conical
bands and is described by a two dimensional analog of the
relativistic Dirac equation.

At the same time, graphene has excellent mechanical
characteristics and is able to sustain huge electric currents
\cite{Novoselov04}. One can attach contacts to submicron graphene
samples and cut out samples of a desired form and size
\cite{Novoselov04,Novosel05}. Applying the electric field one can
vary considerably the electron concentration and have both the
electrons and holes as charge carriers. Due to these properties
the graphene systems look promising for applications in
nanoelectronic devices.

One of the most important directions of research using the
semiconductor heterostructures is fabrication and manipulations
with so called quantum dots that are considered as possible
building blocks for a solid state quantum computer~\cite{Loss98}.
The selection of e.g. $GaAs/AlGaAs$ for this purpose is related to
a possibility of producing electrostatic barriers using weak
electric fields. Changing the field configuration one can change
the form of the quantum dot, its size and other characteristics.
Considering applications of the graphene systems, the fabrication
of the quantum dots looks one the most desirable developments in
the field.

In this paper we discuss a method of making quantum dots in
graphene strips using \emph{electrostatic} gates. At first glance,
the possibility of an electrostatic electron confinement looks
surprising since the total density of the conduction electrons is
huge $n_{e}\approx 4\times 10^{15}cm^{-2}$. However, the striking
feature of the graphene spectrum, namely, the existence of the
degeneracy points makes the local density of the carriers very
sensitive to the electric fields. This opens a way to create
localized states near the zero energy of the $2$-dimensional Dirac
Hamiltonian.

The existence of bound states in a quantum well is one of the
basic features of systems described by the Schr\"{o}dinger
equation.
The situation is different for the Dirac equation,
since chiral relativistic particles may 
penetrate through any high and wide potential barriers. This ideal
penetration~\cite{ando,Cheian06,Katsnel06} means that one cannot
automatically transfer to graphene the experience in fabrication
of quantum dots in $GaAs$ 
using the confinement by
barriers.

Fortunately, one can still localize the charge carriers in the
graphene strip using transversal degrees of freedom of their
motion. Moreover, in most of the examples below the mode ideally
propagating along the strip is prohibited for the strip of a
finite width.

Formation of the quantum dot in a semiconductor wire requires two
tunnelling barriers. Surprisingly, in graphene it is sufficient to
make a single barrier, which may be even simpler from the
experimental point of view. The quasi-bound states exist
\emph{inside} the potential barrier, whose left and right slopes
work as the "tunnelling barriers" for the relativistic electrons.
The width of the energy levels of these quasi-bound states falls
off exponentially with the width of the barrier and can be very
small.

The very existence of the quasi-bound states (resonances) is
independent of the way of the scattering of the electron waves on
the edge of the graphene strip (boundary conditions). However, the
positions and the widths of the individual resonances and
especially the value of the background conductance between the
resonances depends on the type of the boundary. Therefore, an
experimental realization of our setup would allow one to study
properties of the boundary of the real graphene strips.

The electron wave functions in graphene are usually described by a
two component (iso)-spinor $\psi $. Its up- and down- components
correspond to the quantum mechanical amplitudes of finding the
particle on one of the two sublattices of the hexagonal lattice.
In the absence of a magnetic field, the usual electron spin does
not appear in the Hamiltonian and all the electron states have the
extra double degeneracy. The Fermi level of a neutral graphene is
pinned near two corners $\vec{K},\vec{K^{\prime }}$ of the
hexagonal Brillouin zone, which generates two valleys in the
quasiparticle spectrum. The iso-spinor wave function describing
the low energy electron excitations decomposes into a
superposition of two waves oscillating with a very different
wave-vectors $\psi=e^{i\vec{K}\vec{r}}\phi_K
+e^{i\vec{K'}\vec{r}}\phi_{K'}$, where
$\phi_K=(u_{K},v_{K}),\phi_{K'}=(u_{K'},v_{K'})$ are two smooth
enveloping functions. The latter can be found from the
two-dimensional Dirac equation, e.g.
 \begin{equation}
\lbrack c({p_{x}}{\sigma _{x}}+p_{y}\sigma _{y})+V(x)]\phi
_{K}=\varepsilon \phi _{K}.  \label{Dirac}
 \end{equation}
Here $c\approx 10^{8}cm/s$ is the Fermi velocity and
$\vec{p}=-i\hbar \nabla $. We consider the graphene strip of the
width $L$ placed along $x$ axis, $0<y<L$. The smooth potential
$V(x)$ is assumed to be created by an external small size gate
(tip). We consider the simplest case of a parabolic potential
 \begin{equation}
V=-\left( {x}/{x_{0}}\right) ^{2}U/2.  \label{V}
 \end{equation}
Details of the asymptotics of the potential at $|x|\gg x_{0}$ are
not important for our results. The envelope
function~$\phi_{K^{\prime }}$ for the quasiparticle states from
the second valley satisfies the same Eq.~(\ref{Dirac}) with
replaced sublattice indices, i.e. with $\sigma _{y}\rightarrow
-\sigma _{y}$.

Solution of a couple of two-component Dirac
equations~(\ref{Dirac}) in a strip requires a specification of two
boundary conditions at each edge of the strip. We first consider
the "armchair" edge corresponding to the boundary
conditions~\cite{BreyF06} ($y_{1,2}=0,L$)
 \begin{equation}
u_{K}|_{y_{j}}=e^{i2\pi \nu _{j}}u_{K^{\prime }}|_{y_{j}}\ ,\
v_{K}|_{y_{j}}=e^{i2\pi \nu _{j}}v_{K^{\prime }}|_{y_{j}},
\label{armchair}
 \end{equation}
where $i=1,2$ and $\nu _{1}=0$. If the graphene strip contains a
multiple of three rows of the hexagons one obtains $\nu _{2}=0$,
which corresponds to a metal. Other numbers of rows lead to a
semiconducting state with $\nu _{2}=\pm 2/3$.

\begin{figure}
\includegraphics[width=6.2cm]{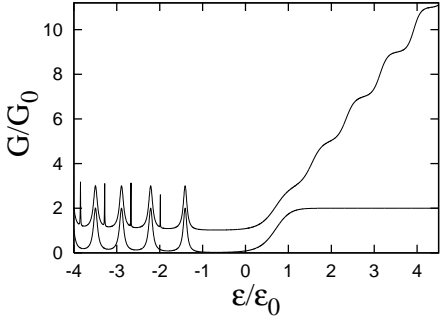}
\vspace{-.3cm} \caption{ Upper curve: Conductance of the graphene
quantum dot as a function of Fermi energy for the metallic
armchair edges for $L=4\protect\xi$, Eq.~(\protect\ref{xi}). Lower
curve: contribution to conductance from the transmission channels
with $p_y=\pm\protect\pi\hbar/L$. All calculations are carried out
for zero temperature, $T=0 $.}
\end{figure}

Eqs.~(\ref{Dirac},\ref{V}) suggest natural units of length and
energy
 \begin{equation}
\xi =\left[ {\hbar c}x_{0}^{2}/U\right] ^{1/3}\ ,\ \varepsilon
_{0}={\hbar c}/{\xi }.  \label{xi}
 \end{equation}
In the experiments \cite{Novoselov04,Novosel05,Zhang05}, a
graphene strip of the width $L\approx 1\mu m$ was separated by a
$.3\mu m$ thick $SiO_{2}$ coating layer from a $n^{+}$ doped $Si$
wafer. We expect that the length scale for the potential $V$, Eq.
(\ref{V}), produced, e.g., by varying the thickness of the
insulator layer, or by local chemical doping, is also
$x_{0}\approx 1\mu m$. Assuming that the characteristic length,
Eq.~(\ref{xi}), is also $\xi \approx 1\mu m$ we estimate
$\varepsilon _{0}\approx U\approx .66\times 10^{-3}eV$.
Making the coordinate dependent potential, Eq.~(\ref{V}), of this
strength looks rather realistic. For example, reaching the carrier
density $n_{s}=p_F^2/\pi\hbar^2\approx 10^{12}cm^{-2}$ would
require a shift of the Fermi energy away from the half-filling by
$\Delta E_{F}=c p_F\approx .12eV$. Even larger carrier densities
in graphene were reported in the
experiments~\cite{Novosel05,Zhang05}.

Solutions of Eqs.~(\ref{Dirac},\ref{V}) for the armchair graphene
strip have a form
 \begin{eqnarray}
&&u_{K}=e^{ip_{y}y/\hbar }(f+g)\ ,\
u_{K^{\prime }}=e^{-ip_{y}y/\hbar }(f+g), \label{uvarmchair} \\
&&v_{K}=e^{ip_{y}y/\hbar }(f-g)\ ,\
v_{K^{\prime}}=e^{-ip_{y}y/\hbar }(f-g). \notag
 \end{eqnarray}
The transverse momentum $p_{y}$ takes the values
 \begin{equation}
p_{y}(n)=(n+\nu _{1}-\nu _{2})\pi \hbar /L\ ,\ n=\cdot \cdot
,-1,0,1,2,\cdots .  \label{py}
 \end{equation}
The Dirac equation~(\ref{Dirac}) is now replaced by
 \begin{eqnarray}
&&\left( -i\hbar {d}/{dx}+V(x)\right) f-icp_{y}g=\varepsilon f,
\label{RDirac} \\
 &&\left( i\hbar {d}/{dx}+V(x)\right)
g+icp_{y}f=\varepsilon g.  \notag
 \end{eqnarray}
These equations decouple from each other and can be solved exactly
provided the momentum component $p_{y}$ perpendicular to the strip
vanishes, $p_{y}=0$,
 \begin{equation}
f=e^{iS},\ \ g=e^{-iS}, \ \ S=\int^x \fr{\varepsilon
-V(x')}{c\hbar}dx' .\label{e1}
 \end{equation}
Eq. (\ref{e1}) is not what we would like to have because it
describes the electron waves propagating \textit{without
reflection} along the strip. This is just the $1d$ solution
considered previously \cite{ando,Cheian06,Katsnel06}.
For $p_{y}\neq 0$, one cannot solve Eqs. (\ref{RDirac}). However,
the {\it exact} asymptotics at $x\rightarrow \pm \infty $ of the
solutions has a simple form (\ref{e1})
 \begin{equation}
f_{-}=e^{iS}\ ,\ g_{-}=re^{-iS}\ ,\ f_{+}=te^{iS}\ ,\ g_{+}=0,
\label{asymptotics}
 \end{equation}
where $r$ and $t$ are two complex numbers, $|r|^{2}+|t|^{2}=1,$
and the subscripts $+,$ $-$ relate to the asymptotics at $\pm
\infty $.

The form of the asymptotics chosen in Eq. (\ref{asymptotics})
corresponds to the electron flux moving from $-\infty $ to
$+\infty ,$ where the coefficient $r$ stands for the reflection
and $t$ for the transmission amplitude. The Landauer formula gives
the conductance $G$ at zero temperature
 \begin{equation}
G=G_{0}\sum |t_{n}|^{2}\ ,G_{0}={2e^{2}}/{h},  \label{Landauer}
 \end{equation}
where $t_{n}=t(p_{y}(n))$. For the metallic armchair edge, the
summation goes over $n=0,\pm 1,\pm 2,\cdots $, and
$|t_{n}|^{2}=|t_{-n}|^{2}$. The factor $2$ in $G_{0}$ accounts for
the electron spin.

Analytical calculation of the transmission coefficients $t_{n}$ is
possible only for $|\varepsilon |\gg \varepsilon _{0}$. Fig.~1
shows the dependence of the conductance on the Fermi energy
$\varepsilon $ calculated numerically for $L=4\xi $~for the
metallic armchair strip. At $\varepsilon >0$ the conductance
increases linearly with clearly visible steps $\Delta G\approx
2G_{0}$ corresponding to the opening of new
channels~\cite{Been91}. At negative Fermi energies one can see a
series of pronounced resonances. The resonances appear for all
nonzero values of the transverse momentum $p_{y}\neq 0$~(\ref{py})
but only those corresponding to $|n|=1$ and $|n|=2$ are resolved
in the figure.

It is easy to understand the reason for the appearance of the
(quasi)bound states in graphene. Solutions of the equation
$V(x)=\varepsilon $ divide the strip into regions with electron or
hole type of carriers. The lines separating these regions serve as
tunnelling barriers for all but normal
trajectories~\cite{Cheian06} and this leads to the confinement.

\begin{figure}
\includegraphics[width=6.2cm]{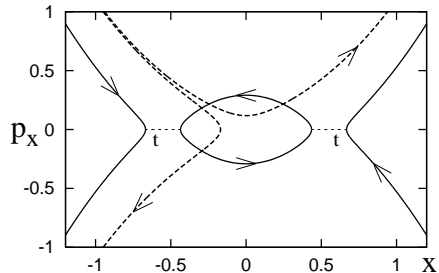}
\vspace{-.3cm} \caption{Examples of trajectories described in the
text drawn on the $x,p_x$ plane (arbitrary units). Solid lines
show the trajectories with $\eps<0$ either bouncing inside the
barrier, or reflected by it from the left/right. Tunnelling events
between the bounded and unbounded trajectories are shown
schematically (t). Thick dashed lines show the trajectories with
$\eps>0$ either transmitted for $|p_y|<\eps/c$ (open channels) or
reflected for $|p_y|>\eps/c$ (closed channels). }
\end{figure}

The (semi)classical dynamics of the massless Dirac fermions,
Eq.~(\ref{Dirac}), is given by the effective Hamiltonian (see
examples of classical trajectories in Fig.~2)
 \begin{equation}
H_{\mathrm{eff}}=\varepsilon =\pm
c\sqrt{p_{x}^{2}+p_{y}^{2}}+V(x). \label{Heff}
 \end{equation}
For the $(+)$ sign particles may either fly freely above the
barrier $V(x)$, or start at the infinity and then be reflected
from the barrier. The $(-)$ sign in $H_{\mathrm{eff}}$ corresponds
to the hole solutions of the Dirac equation, whose trajectories
bounce inside the barrier for our choice of $V(x)$, Eq.~(\ref{V}).
For a given value of the transverse momentum $p_{y}\neq 0$ four
classical turning points where the trajectory changes the
direction of the propagation along the strip ($p_{x}=0$) are
 \begin{equation}
\frac{x_{\mathrm{out}_{\pm }}}{x_{0}}=\pm
\sqrt{2\frac{c|p_{y}|-\varepsilon}{U}}\ ,\
\frac{x_{\mathrm{in}_{\pm }}}{x_{0}}=\pm
\sqrt{2\frac{-c|p_{y}|-\varepsilon }{U}}.  \label{turning}
 \end{equation}
The electron coming from the infinity is reflected by one of the
outer turning points $x_{\mathrm{out}_{\pm }}$ if
$c|p_y|>\varepsilon$. Thus changing the energy $\varepsilon $ one
changes the number of open channels, which leads to the
conductance quantization for positive $\varepsilon $ with the
smoothed conductance $G\approx 2G_{0}L\varepsilon /(\hbar \pi c)$.

Finite (hole)trajectories bouncing between the two inner turning
points $x_{\mathrm{in}_{\pm }}$ give rise to the quasi-stationary
states. These trajectories appear at $\varepsilon <-c|p_{y}|$ and
the position of the $N$-th resonance $\varepsilon _{N}$ may be
found from the quasiclassical quantization rule
 \begin{equation}
\int_{x_{\mathrm{in}_{-}}}^{x_{\mathrm{in}_{+}}}\sqrt{(\varepsilon
_{N}-V(x))^{2}-c^{2}p_{y}^{2}}\frac{dx}{c}=\pi \hbar (N+\frac{1}{2}).
\label{BSommerfeld}
 \end{equation}
The resonance acquires a finite width due to quantum tunnelling
between $x_{\mathrm{in}}$ and $x_{\mathrm{out}}$. We can estimate
the width using a result of Ref.~\cite{Cheian06}, where the
transmission probability through a linear potential was obtained
in the form $w=\exp (-\pi cp_{y}^{2}/\hbar F)$, where $F=|dU/dx|$
is a slope of the potential. This result can be used in our case
provided $c|p_{y}|\ll |\varepsilon _{N}|$. Since in this case the
interval $\Delta t$ between the reflections at the points
$x_{\mathrm{in}_{-}},$ $x_{\mathrm{in}_{+}}$ equals $\Delta
t=2(x_{0}/c)\sqrt{-2\varepsilon _{N}/U}$, we find the width
 \begin{equation}
\Gamma _{N}=\frac{\hbar }{\Delta t}w=\frac{\hbar
v_{0}}{2x_{0}}\sqrt{\frac{U}{-2\varepsilon _{N}}}\exp \left(
-\frac{\pi cp_{y}^{2}x_{0}}{\hbar \sqrt{-2\varepsilon
_{N}U}}\right) . \label{Gamma}
 \end{equation}
Increasing the characteristic length of the potential $x_{0}$ we
can get an extremely narrow levels and long time of the
confinement of the electrons in such a quantum dot.

The above results have been obtained for the graphene strip with
the metallic armchair edges ($\nu _{2}=0$ in
Eq.~(\ref{armchair})). Specific for such edges is the existence of
the channel with $p_{y}=0$ providing the perfect transmission at
any energy $|t_{0}(\varepsilon )|^{2}\equiv 1$. Below we describe
the conductance behavior for few other kinds of the edge.

Fig.~3 shows the conductance of the semiconductor armchair
graphene strip ($\nu _{2}=\pm 2/3$ in Eq. (\ref{armchair})) as
compared to the conductance of the metallic one, both found from
Eqs.~(\ref{RDirac}-\ref{Landauer}). Several striking differences
between the two kinds of the edges are clearly seen in the figure.
First, since there is no channel with $p_{y}=0$, the background
conductance around the resonances at $\varepsilon <0$ vanishes for
the semiconductor strip, $G\ll G_{0}$. In the metallic case the
averaged conductance at $\varepsilon <0$ is $G\approx G_{0}$.
Second, the height of the conductance steps at $\varepsilon >0$ is
$\Delta G=2G_{0}$ for the metallic graphene strip and $\Delta
G=G_{0}$ for the semiconductor one. Third, the length of the
conductance plateaus is constant in the metallic case. On the
contrary, the conductance steps in the semiconducting strip have
alternating short and long plateaus with $\Delta
\varepsilon_{2}\approx 2\Delta \varepsilon _{1}$. In the
experiment, one can expect that the metallic and semiconductor
strips will be produced in a proportion~$1:2$.

A way to define the boundary of a Dirac billiard was proposed many
years ago in Ref.~\cite{BerryMon87} by introducing an infinite
mass for quasiparticle behind the boundary. Ref.~\cite{Tworzyd06}
suggested that in graphene this boundary would correspond to the
transverse confinement of carriers by lattice straining. The two
($K,K^{\prime }$) valleys in this case are decoupled from each
other and one has~\cite{endnoteedge}
 \begin{eqnarray}
u_{K}(0)=v_{K}(0) &,&u_{K}(L)=-v_{K}(L),  \label{straining} \\
u_{K}+v_{K}=f(x)\cos \frac{p_{y}y}{\hbar } &,&u_{K}-v_{K}=g(x)\sin
\frac{p_{y}y}{\hbar },  \notag
 \end{eqnarray}
where the functions $f$ and $g$ are the solutions of
Eq.~(\ref{RDirac}). The boundary conditions~(\ref{straining}) are
satisfied for
$p_{y}=(n+{1}/{2})\pi {\hbar }/{L}$, $n=0,1,2,\cdots$.
Each solution, Eq. (\ref{straining}), is fourfold degenerate.
Since $p_{y}\neq 0$, the conductance around the resonances is
zero, $G\ll G_{0}$. The curve $G(\varepsilon )$ is now very
similar to what we have found for the metallic armchair edges
(Fig.~1) provided the picture is shifted vertically by $-G_{0}$.

\begin{figure}
\includegraphics[width=6.2cm]{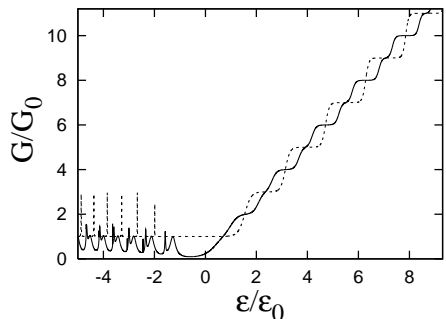}
\vspace{-.3cm} \caption{Conductance of the graphene quantum dot
for the semiconductor armchair edge (solid line) compared to the
metallic armchair edge (dashed line) for $L=2\protect\xi$
(\protect\ref{xi}). The non-resonant conductance at
$\protect\varepsilon<0$ is $G\approx G_0$ for the metallic strip
and $G\approx 0$ for the semiconducting one. At
$\protect\varepsilon>0$ the conductance steps in the semiconductor
case are two times smaller, $\Delta G\approx G_0$. Alternating
series of short
and long
steps are clearly seen for the semiconductor strip.}
\end{figure}

Another widely considered type of the edge in graphene is the
zigzag edge. Since in the case of the zigzag boundary the edges of
the strip belong to the different sublattices, the components $u$
and $v$ of the envelope function vanish at the opposite sides of
the strip
 \begin{equation}
u_{K}(0),u_{K^{\prime }}(0)=0\ ,\ v_{K}(L),v_{K^{\prime }}(L)=0.
\label{zigzag}
 \end{equation}
In addition to the solutions described by the Dirac equation
(\ref{Dirac}), the zigzag edge supports a band of zero energy edge
states~\cite{Fujita96,Ryu02}. The (unknown) conductance $\sim
G_{0}$ due to the edge states should be added to the bulk
conductance~(\ref{Landauer}). Except for this edge states
contribution we do not expect significant differences between the
conductance of the strips with the zigzag edges  and those
confined by the lattice straining~\cite{zigzag}.

In Figs.~1,3 the heights of the resonances are determined by the
level of degeneracy (two- or fourfold) of the bound states of the
\textit{non-interacting} electrons. In the experiment the shape of
the resonances will be governed by the electron
\textit{interaction} via the Coulomb blockade
effect~\cite{CB,Bunch05}.
Since the widths of the resonances corresponding to large values
of the transverse momentum~(\ref{Gamma}) become exponentially
small, the multiple charging and repopulation of a broad level
introduced in~Ref.~\cite{Sil00} may occur here.

To conclude, we considered a possibility of localizing charge
carriers in a graphene strip by applying an external electrostatic
potential. Such a quantum dot can be fabricated using a parabolic
potential with a single maximum(minimum). Depending on the
position of the Fermi energy, such a device can serve as either a
quantum dot, or a quantum point contact. The two regimes
correspond to either the resonance conductance or the quantized
(step-like) one. An experimental realization of our findings would
open a way to investigate in graphene the reach physics of
individually prepared quantum dots.

This work was supported by the SFB TR 12. Discussions with
A.F.~Volkov are greatly appreciated.


\begin{thebibliography}{99}
\bibitem{endnoteedge} Reflection from the graphene boundary should
naturally respect the conservation of the number of particles. One
easily shows that on the class of boundaries which do not allow
transitions between valleys $(K,K')$ the only two posisibilities
to conserve the number of particles for any incident angle are
given by Eqs.~(\ref{straining}) and (\ref{zigzag}).
\bibitem{zigzag} Eq.~(\ref{Dirac}) for zigzag edges
can be solved
analytically in the limit of a smooth potential, $\xi \gg L$. Then
$
u_{K,K^{\prime }}=\sqrt{\rho (x)}e^{i\int p_{x}{dx}/{\hbar }}\sin
{p_{y}y}/{\hbar }=\pm v_{K,K^{\prime }}(L-y),
$
where both $p_{x}(x)$ and $p_{y}(x)$ depend (slowly) on the
coordinate $x$ with the constraint $p_{x}\sin ({p_{y}L}/{\hbar
})+p_{y}\cos ({p_{y}L}/{\hbar })=0$. A conduction channel opens
when a new solution appears at the maximum, $x=0$, of the
potential $V(x)$. At the pinch-off one has $p_{x}(0)=0$ and this
leads to $p_{y}(x=0)=(n+{1}/{2})\pi {\hbar }/{L}\ ,\
n=0,1,2,\cdots$.
\end{thebibliography}
\end{document}